\documentclass[conference,a4paper]{IEEEtran}
\pdfoutput=1
\ifCLASSINFOpdf
   \usepackage[pdftex]{graphicx}
\else
\fi

\usepackage[table]{xcolor}

\usepackage[font=footnotesize]{subfig}
\usepackage{xfrac}

\definecolor{Gray}{gray}{0.85}
\begin{document}
%
\title{Neural Network based Inter bi-prediction Blending}

\author{
\IEEEauthorblockN{Franck Galpin\IEEEauthorrefmark{2},
Philippe Bordes\IEEEauthorrefmark{2},
Thierry Dumas\IEEEauthorrefmark{2},
Pavel Nikitin\IEEEauthorrefmark{2},
Fabrice Le Leannec\IEEEauthorrefmark{2}
}
\IEEEauthorblockA{\IEEEauthorrefmark{2}InterDigital Inc.\\
firstname.lastname@interdigital.com}
}

%


\maketitle

\begin{abstract}
This paper presents a learning-based method to improve bi-prediction in video coding. In conventional video coding solutions, the motion compensation of blocks from already decoded reference pictures stands out as the principal tool used to predict the current frame. Especially,  the bi-prediction, in which a block is obtained by averaging two different motion-compensated prediction blocks, significantly improves the final temporal prediction accuracy.

In this context, we introduce a simple neural network that further improves the blending operation. A complexity balance, both in terms of network size and encoder mode selection, is carried out. Extensive tests on top of the recently standardized VVC codec are performed and show a BD-rate improvement of -1.4\% in random access configuration for a network size of fewer than 10k parameters. We also propose a simple CPU-based implementation and direct network quantization to assess the complexity/gains tradeoff in a conventional codec framework.

\end{abstract}


%
\IEEEpeerreviewmaketitle

\section{Introduction}
In conventional hybrid video codec, inter frames are used to increase the compression performance drastically by allowing prediction of a given block from a motion-compensated version of the block in one or several reference frames already available at the decoder. In recent new codec generation such as VVC \cite{vvc}, the model used for motion compensation has been further improved to include affine motion model, temporal sub-block prediction (SBTMVP), or sub-block based decoder motion refinement (DMVR). For each block of an inter frame, the prediction can be computed using one or two different motion compensated blocks, the latter being called bi-prediction mode. Bi-prediction blends together two different versions of the same block, usually coming from two different reference frames. The blending of the two blocks allows mitigating errors due to coding artifacts, illumination changes, small local motions inside the block.

In this paper, we investigate the improvement of the blending stage of the bi-prediction process where two predictions are blended together to form the final prediction. We show that a Neural Network based approach allows us to further improve already existing methods and study the complexity tradeoff related to this kind of approach.

The rest of this paper is organized as follows. Related work is reviewed in Sec. \ref{relatedwork}. The method is described in detail in Sec. \ref{method}. Experimental results are discussed in Sec. \ref{results}. Finally, concluding remarks are given in Sec. \ref{ccl}.

\section{Related work}\label{relatedwork}
In recent codecs such as VVC, the bi-prediction process has been improved by the introduction of several new tools. First, a weighted prediction of the two temporal predictions with a weight signaled per block has been introduced. The tool, called BCW (bi-prediction at CU-level weights \cite{vvc}), uses a set of fixed weights in $[ -\sfrac{1}{4},  \sfrac{3}{8}, \sfrac{1}{2}, \sfrac{5}{8}, \sfrac{5}{4}]$ (the sum of the weights of the two predictions being 1.0). 
Another improvement of the bi-prediction introduces a correction term on top of the bi-prediction by computing an offset derived from the optical flow equation between the two prediction blocks \cite{bdof}. The process called BDOF (bidirectional optical flow) assumes that the two blocks are referring to frames at equal distance in the past and the future of the current block, and that the underlying motion of the pixels in the block has a constant speed between the two blocks. By simply computing the spatial and temporal gradients for each pixel, the correction term can then be deduced from the optical flow equation.
Another improvement of the bi-prediction blending was introduced by the GPM (Geometric Partitioning Mode \cite{vvc}) where the two predictions are blended differently for each sub-block inside the block. The weight of each prediction is adapted depending on the distance to a boundary-crossing the block, where the weights are typically $[\sfrac{1}{2}, \sfrac{1}{2}]$ in the subblocks containing the boundary.
Finally, another type of improvement of the prediction is the CIIP (Combined inter-/intra-picture prediction \cite{vvc}) tool, which combines an intra prediction of the block with an inter prediction.

Following the success of Deep-Learning based methods in many areas, some new tools were also developed in the area of video compression. One of the most popular is the application of image restoration to the post-filters of the video codec \cite{vrcnn}. Recently, applying the ideas of image inpainting, intra prediction tools were also successfully designed \cite{intra}.

Apart from the traditional approach in video coding, recent progress using Neural Networks based approaches show significant improvements in temporal prediction with the introduction of deep frame interpolation tools. In these approaches, two or more frames are used to generate a frame between these two frames (interpolation mode) or after the two frames (extrapolation mode). One type of approach consists of explicitly using a motion field between the two frames, for example, in \cite{deepvoxel} and then use this motion field to create the interpolated frame. Note that these approaches are still unsupervised since the final loss can be expressed using only the interpolated frame and the real frame.
Another type of approach is to directly optimize the reconstruction of the interpolated frame as in \cite{sepconv}. In this case, the network might have more "freedom" to create the interpolated frame, such as performing illumination correction, occlusions aware interpolation.

The methods above have then been naturally used in the context of video coding where the frame interpolation problem is formulated as a frame prediction problem. For example, in \cite{deepframe} or \cite{deepframe2} a new temporal prediction candidate is created from already decoded frames. In these approaches, the neural network is run both at encoder and decoder to create the prediction, avoiding the transmission of motion information. However, the complexity increases significantly, especially on the decoder side. It would also require running the network for each block to decode, or alternatively for the whole frame once. Moreover, handling very large motion is still challenging with these approaches and the best results are achieved for low-delay configuration where encoding is performed on a frame to frame basis.

More advanced methods, for example, dealing with large or complex motion and producing perceptually better interpolated frames like in \cite{slowmo} have also been developed but might not be suitable in the context of video coding due to their high complexity and the use of a perceptual loss in the rate-distortion optimization loop.


\section{Proposed method}\label{method}

\subsection{Description}
\begin{figure}[htp]
\centering
\includegraphics[width=8cm]{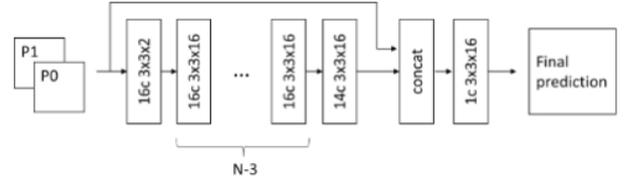}
\caption{Network geometry overview.}
\label{fig_nn}
\end{figure}

The goal of the new tool is to perform the blending of two temporal predictions. Using a neural network, a more complex, non linear, blending is expected, taking into account both local motion, illumination change or other non obvious temporal changes between frames. For this purpose, we designed a relatively small, fully convolutional neural network that takes as input the two predictions and outputs the final prediction. Figure \ref{fig_nn} shows the general layout of the network:
\begin{itemize}
\item A first convolutional layer, parametrized by 16 3x3 convolutions, takes as input the two prediction blocks.
\item N-3 convolutional layers of 16 3x3 convolutions.
\item A convolutional layer of 14 3x3 convolutions.
\item A concatenation between the two inputs and the output of the above layer. Compared to just adding a residual to the inputs, it allows richer combination of the inputs and the correction term thanks to the last convolution layer.
\item A final convolution layer of 1 convolution layer.
\end{itemize}
All layers, but the last one uses a ReLU activation layer which allows simple implementation. The last one uses a clip after the convolution. The input blocks are enlarged with a border of N pixels on each side.
The layer before the concatenation has a depth of 14 in order to get a layer of depth 16 after concatenation with the two inputs. This allows a more friendly implementation when using SIMD code on CPU (see sec. \ref{inference}). 

\subsection{Tradeoffs}

\subsubsection{Conditions of application}
The tool is applied without the need of additional signaling in the bitstream but some conditions are set in order to decide if the tool is applied or not. When not applied, the default blending is applied.
As the role of the network is somewhat similar to the BDOF tool and the 2 modes are not compatible since they both change the final prediction from input predictions without signaling, BDOF is deactivated in our tests. The network is currently trained and used only for the luma component; regular blending is applied on chroma components.
In order to test some tradeoffs, 3 conditions sets are tested as depicted in Table \ref{conditions}. Note that the worst case at the decoder is not changed, but the encoder will be faster when the conditions are stricter. 
\begin{table}[htp]
\begin{center}
\caption{Mode conditions}
\label{conditions}
\begin{tabular}{|c|l|}
\hline
 \rowcolor{Gray}\textbf{Mode} & \textbf{Conditions} \\
\hline
default & a) CU is not affine \\
       &  b) CU does not use CIIP\\
& c) CU does not use BCW \\
 & d) POC of reference blocks symmetrically in past and future\\
&    e) CU does not use SMVD \\
\hline
fast & same as default \\
 & + block width and height greater than 8 \\
 \hline
slow & same as default \\
  & without condition d) and e) \\
  \hline
\end{tabular}
\end{center}
\end{table}

\subsubsection{Network size}
To assess the impact of the network size on performance/complexity tradeoffs, two versions of the network are proposed with N=5 or N=6. One reason to limit the complexity, besides the increase of decoder complexity, is also because the tool will be used inside the RDO (Rate-distortion Optimization) loop on the encoder side and the network will be called extensively (typically several 10k times per CTU for the current VTM implementation of VVC). 
Table \ref{complexity} shows the associated complexity and memory usage for each network. The MAC/pixel reflects the computational complexity of the network and the peak memory usage of the memory bandwidth needed for the biggest layer, as advise in \cite{ahg11}.


\begin{table}[htp]
\begin{center}
\caption{Information at inference stage}
\label{complexity}
\begin{tabular}{|c|c|c|}
\hline
 \rowcolor{Gray}\textbf{Properties} & \textbf{small (N=5)} & \textbf{medium (N=6)} \\
\hline
Runtime & CPU single thread & CPU single thread \\
\hline
Framework & custom implementation& custom implementation \\
\hline
Number of parameters & 7k & 9k \\
\hline
Parameter precision (Bits) & 16 (I) & 16 (I) \\
\hline
Memory Parameter &  13 kB  & 17.8 kB \\
\hline
Complexity (max) & 11.3 kMACs/pix &  16.6 kMACs/pix\\
\hline
Peak Memory Usage & 95kB  & 110kB \\
\hline
\end{tabular}\end{center}
\end{table}

\subsection{Inference}\label{inference}
A custom CPU-only based implementation is used for inference.
It has several advantages: first, it allows a realistic estimate of the added complexity in the same conditions as other tools already present in the codec (as opposed to using a GPU-based or special framework for inference).
The framework is a single thread and uses naïve SIMD optimization for some operations (typically the convolutions) in order to follow the already established CTC of JVET (Common Test Conditions see \cite{ctc}). 
In order to reduce the complexity and allow reproducible results, the framework uses integer operations only. The network parameters are quantized on 16 bits, internal computation is done on 32 bits and intermediate values are quantized on 16 bits.

As the network is relatively small and the quantization uses 16 bits values, direct quantization of each layer is performed. In practice, each layer of the network uses a fixed-point representation of the float values, with the quantizer for each layer being chosen optimally to minimize the error on the final output compared to the float version. With this method, no re-training is necessary to obtain the same accuracy as the float version of the network. The Table \ref{perftable} shows the performance of the framework compared to TensorFlow \footnote{Tensorflow-cpu 2.3, oneDNN, ext: AVX512F/FMA, single-thread mode}. Cold start is measured as the first call to the network inference, while warm start measures the average inference time of successive inferences.


\begin{table}[htp]
\begin{center}
\caption{Inference performance for a 32x32 patch.}

\begin{tabular}{|c|c|c|}
\hline
 \rowcolor{Gray}\textbf{Inference} & \textbf{cold start (ms)} & \textbf{warm start (ms)} \\
\hline
Tensorflow CPU - single thread & 88.4  & 2.3\\
\hline
Custom int16 & 1.3 & 1.0\\
\hline
\end{tabular}
\end{center}
\end{table}\label{perftable}

The current implementation in the encoder is straightforward and no heuristic is used in the RDO. Specifically, pessimistic block prefetching is used, slowing down the encoding process.

\subsection{Training details}
Tensorflow 2.0 is used as the training platform. The BVI-DVC\cite{bvi} and UVG\cite{uvg} datasets are used to train the network. The network information in the training stage is provided in Table \ref{training}. The pair of blocks are extracted using blocks used in BDOF for a subset of each sequence in the dataset. The total dataset contains $\sim$50M of blocks. 
As depicted in the table, we use the SATD (Sum of Absolute Transform Difference) as the error function in the loss, using the Hadamard matrix for the transformation. This function is widely used in video codec to get a better estimate of the prediction error as it does not only take into account the pixel to pixel error (as it is typically done by a $L_1$ or $L_2$ norm), but also try to assess the cost of the residual error coding, which is ultimately what we want to minimize.

\begin{table}[htp]
\begin{center}
\caption{Network Information for NN-based Video Coding Tool Testing in Training Stage}
\label{training}
\begin{tabular}{|c|c|}
\hline
 \rowcolor{Gray}\textbf{Properties} & \textbf{Value} \\
\hline
GPU Type & Tesla-P100-16GB \\
\hline
Framework & Tensorflow 2.x \\
\hline
Number of epoch & 10 \\
\hline
Batch size & 256 \\
\hline
Training time & $\sim 100h$\\
\hline
Dataset creation & VTM-11.0, QP {22, 27, 32, 37, 42} \\
\hline
Patch size & 16x16x2 \\
\hline
Learning rate & 1e-4 \\
\hline
Optimizer & ADAM \\
\hline
Loss function & SATD \\
\hline
\end{tabular}
\end{center}
\end{table}

\section{Results}\label{results}

For all results but the last one, the anchor is VTM-11.0. The CTC \cite{ctc} are used for sequences, configuration and BD-rate computation but we used 5 QPs (22, 27, 32, 37, 42) instead of 4 for computation. BD-rate gains at constant PSNR are given for Y, U and V and encoding and decoding time changes compared to anchors are given in EncT and DecT columns.

\subsection{Network size comparison}
We first compare the effect of the network size in Tables \ref{nn_small} and \ref{nn_medium}, both using the default configuration in Random Access configuration. A bigger network adds almost 0.5\% gains with a decoder complexity increase of almost 40\%.

\begin{table}[htp]
\centering
\caption{Small - default conditions - RA configuration}
\begin{tabular}{|c|c|c|c|c|c|}
\hline
 \rowcolor{Gray} 
                & Y-PSNR  & U-PSNR  & V-PSNR  & EncT  & DecT   \\ \hline
\cellcolor{Gray}Class A1         & -0.26\% & -0.05\% & 0.05\%  & 481\% & 579\%  \\ \hline
\cellcolor{Gray}Class A2         & -0.77\% & -0.30\% & -0.21\% & 554\% & 832\%  \\ \hline
\cellcolor{Gray}Class B          & -0.56\% & -0.28\% & -0.20\% & 553\% & 751\%  \\ \hline
\cellcolor{Gray}Class C          & -0.76\% & -0.39\% & -0.32\% & 488\% & 825\%  \\ \hline
\cellcolor{Gray}Class D          & -1.51\% & -1.04\% & -0.77\% & 500\% & 1165\% \\ \hline
\cellcolor{Gray}\textbf{Overall} & \textbf{-0.79\%} & \textbf{-0.43\%} & \textbf{-0.31\%} & \textbf{521\%} & \textbf{812\%}  \\ \hline
\end{tabular}\label{nn_small}
\end{table}

\begin{table}[htp]
\centering
\caption{Medium - default conditions - RA configuration}
\begin{tabular}{|c|c|c|c|c|c|}
\hline
    \rowcolor{Gray}               & Y-PSNR  & U-PSNR  & V-PSNR  & \multicolumn{1}{l|}{EncT} & \multicolumn{1}{l|}{DecT} \\ \hline
\cellcolor{Gray}Class A1         & -0.42\% & -0.15\% & 0.04\%  & 648\%                     & 752\%                     \\ \hline
\cellcolor{Gray}Class A2         & -1.06\% & -0.37\% & -0.31\% & 757\%                     & 1082\%                    \\ \hline
\cellcolor{Gray}Class B          & -0.83\% & -0.40\% & -0.35\% & 731\%                     & 1033\%                    \\ \hline
\cellcolor{Gray}Class C          & -1.24\% & -0.46\% & -0.41\% & 645\%                     & 1168\%                    \\ \hline
\cellcolor{Gray}Class D          & -2.68\% & -1.45\% & -1.10\% & 654\%                     & 1624\%                    \\ \hline
\cellcolor{Gray}\textbf{Overall} & \textbf{-1.28\%} & \textbf{-0.59\%} & \textbf{-0.45\%} & \textbf{694\%}                     & \textbf{1109\%}                    \\ \hline
\end{tabular}\label{nn_medium}
\end{table}

\subsection{Conditions comparisons}
We then compare the effect of the conditions of application in Tables \ref{nnfast} and \ref{nnslow}, both using the medium network in the configuration in Random Access configuration. With the slow configuration, up to 2.9\% can be reached in RA configuration for class D, and 1.4\% on average on all classes. On the encoder side, we notice that a bigger network with the fast configuration gives a better tradeoff than a smaller network in the default configuration. However, on the decoder side, complexity does increase by using the bigger network.
\begin{table}[htp]
\centering
\caption{Medium - fast conditions - RA configuration}
\begin{tabular}{|c|c|c|c|c|c|}
\hline
   \rowcolor{Gray}                & Y-PSNR  & U-PSNR  & V-PSNR  & \multicolumn{1}{l|}{EncT} & \multicolumn{1}{l|}{DecT} \\ \hline
\cellcolor{Gray}Class A1         & -0.42\% & -0.12\% & 0.01\%  & 648\%                     & 752\%                     \\ \hline
\cellcolor{Gray}Class A2         & -0.95\% & -0.24\% & -0.20\% & 757\%                     & 1082\%                    \\ \hline
\cellcolor{Gray}Class B          & -0.71\% & -0.27\% & -0.18\% & 731\%                     & 1033\%                    \\ \hline
\cellcolor{Gray}Class C          & -0.97\% & -0.38\% & -0.28\% & 645\%                     & 1168\%                    \\ \hline
\cellcolor{Gray}Class D          & -2.11\% & -1.30\% & -0.93\% & 654\%                     & 1624\%                    \\ \hline
\cellcolor{Gray}\textbf{Overall} & \textbf{-1.05\%} & \textbf{-0.48\%} & \textbf{-0.33\%} & \textbf{531\%}                     & \textbf{1017\%}                    \\ \hline
\end{tabular}\label{nnfast}
\end{table}

\begin{table}[htp]
\centering
\caption{Medium - slow conditions - RA configuration}
\begin{tabular}{|c|c|c|c|c|c|}
\hline
      \rowcolor{Gray}             & Y-PSNR  & U-PSNR  & V-PSNR  & \multicolumn{1}{l|}{EncT} & \multicolumn{1}{l|}{DecT} \\ \hline
\cellcolor{Gray}Class A1         & -0.46\% & -0.14\% & -0.01\% & 732\%                     & 911\%                     \\ \hline
\cellcolor{Gray}Class A2         & -1.16\% & -0.40\% & -0.28\% & 835\%                     & 1221\%                    \\ \hline
\cellcolor{Gray}Class B          & -0.88\% & -0.29\% & -0.31\% & 793\%                     & 1106\%                    \\ \hline
\cellcolor{Gray}Class C          & -1.40\% & -0.50\% & -0.55\% & 692\%                     & 1252\%                    \\ \hline
\cellcolor{Gray}Class D          & -2.89\% & -1.38\% & -1.27\% & 691\%                     & 1643\%                    \\ \hline
\cellcolor{Gray}\textbf{Overall} & \textbf{-1.39\%} & \textbf{-0.56\%} & \textbf{-0.51\%} & \textbf{754\%}                     & \textbf{1211\%}                    \\ \hline
\end{tabular}\label{nnslow}
\end{table}

\subsection{LDB configuration}
Another interesting point is that the tool still achieves some gains in LDB configuration (see Table \ref{nn_ldb}), whereas in this configuration, reference frames used for bi-prediction are not symmetrically placed. Typically, in this mode, the BDOF tool gains are quite low compare to this tool (for this reason, the tool is deactivated in LDB configuration).

\begin{table}[htp]
\centering
\caption{Medium - normal conditions  - LDB configuration}
\begin{tabular}{|c|c|c|c|c|c|}
\hline
    \rowcolor{Gray}
                   & Y       & U       & V      & EncT  & DecT  \\ \hline
\cellcolor{Gray}Class B          & -0.36\% & 0.13\%  & 0.30\% & 575\% & 895\% \\ \hline
\cellcolor{Gray}Class C          & -0.34\% & 0.05\%  & 0.09\% & 433\% & 671\% \\ \hline
\cellcolor{Gray}Class D          & -0.85\% & 0.07\%  & 0.00\% & 392\% & 582\% \\ \hline
\cellcolor{Gray}Class E          & -0.91\% & -0.45\% & 0.39\% & 386\% & 431\% \\ \hline
\cellcolor{Gray}\textbf{Overall} & \textbf{-0.58\%} & \textbf{-0.01\%} & \textbf{0.19\%} & \textbf{452\%} & \textbf{652\%} \\ \hline
\end{tabular}\label{nn_ldb}
\end{table}

\subsection{Raw gains}
Finally, another interesting result comes from the comparison of the tool with a VTM-11.0 anchor without the BDOF tool, as the proposed tool is replacing this mode. In table \ref{nn_bdof}, we see that raw gains of the tool are about 2.2\%.

\begin{table}[htp]
\centering
\caption{Medium - default conditions  - RA configuration -- BDOF off in anchor}
\begin{tabular}{|l|l|l|l|l|l|}
\hline
   \rowcolor{Gray}        & Y-PSNR  & U-PSNR  & V-PSNR  & EncT  & DecT   \\ \hline
\cellcolor{Gray}Class A1 & -0.89\% & -0.41\% & -0.08\% & 677\% & 755\%  \\ \hline
\cellcolor{Gray}Class A2 & -2.18\% & -0.80\% & -0.73\% & 777\% & 1080\% \\ \hline
\cellcolor{Gray}Class B  & -1.65\% & -0.83\% & -0.78\% & 758\% & 1113\% \\ \hline
\cellcolor{Gray}Class C  & -2.09\% & -1.02\% & -0.95\% & 662\% & 1283\% \\ \hline
\cellcolor{Gray}Class D  & -4.15\% & -2.22\% & -1.99\% & 666\% & 1784\% \\ \hline
\cellcolor{Gray}\textbf{Overall}  & \textbf{-2.23\%} & \textbf{-1.09\%} & \textbf{-0.95\%} & \textbf{715\%} & \textbf{1177\%} \\ \hline
\end{tabular}\label{nn_bdof}
\end{table}

\section{Discussions}\label{ccl}

\subsection{Visual analysis}

\begin{figure}[htp]
\centering
\includegraphics[width=8cm]{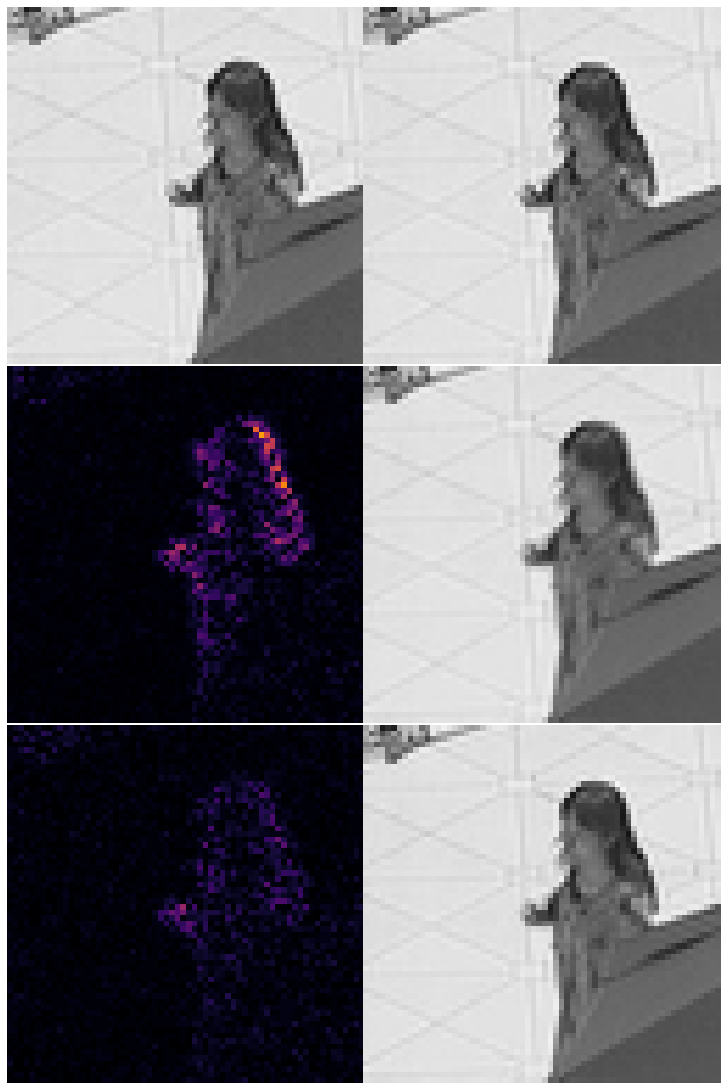}
\caption{Example of blending on BQSquare sequence.}
\label{fig_sim}
\end{figure}
In Figure \ref{fig_sim}, we show a simulation of the network by using the first 3 consecutive frames from the sequence BQSquare (see \cite{ctc}). Prediction 0 and 1 are shown on the top row. The default blending (averaging of the two predictions) is shown in the middle row. We note that the edges are blurry on moving objects and the final PSNR with the original frame is 35.4dB. In the last row, the proposed blending shows sharp edges and a PSNR of 39.5dB. On the left side, difference images also show the error reduction on edges.

\section{Conclusion}
We first show that good performance results can be achieved by introducing neural network-based methods at the blending stage of the bi-prediction on top of the newest video codec as VVC. It is shown that the bigger network has a better trade-off of gain/MAC, typically  $\sim$17kMAC/pix per 1\% BD-rate gains. In practice, the tool allows more than double the gains of BDOF. 
The proposed tool also allows gains of -0.58\% in LDB configuration, whereas BDOF had smaller gains when applied in LDB configuration.

However, even if the trade-off gains/MAC is higher compared to other Neural Network based approaches (intra prediction, post-filters etc.), the overall complexity remains quite high compared to non neural network based approaches, and further investigation is needed to improve the overall complexity/gains trade-off.






%
\bibliographystyle{IEEEtran}
\bibliography{IEEEabrv,mybib}

\end{document}